# Ice Ic without stacking disorder by evacuating hydrogen from hydrogen hydrate

Kazuki Komatsu[1*], Shinichi Machida[2], Fumiya Noritake[3,4], Takanori Hattori[5], Asami Sano-Furukawa[5], Ryo Yamane[1], Keishiro Yamashita[1] & Hiroyuki Kagi[1]

[1] Geochemical Research Center, Graduate School of Science, The University of Tokyo, 7-3-1 Hongo, Bunkyo-ku, Tokyo 113-0033, Japan

[2] Neutron Science and Technology Center, CROSS, 162-1 Shirakata, Tokai, Naka, Ibaraki 319-1106, Japan

[3] Graduate Faculty of Interdisciplinary Research, University of Yamanashi, 4-3-11 Takeda, Kofu, Yamanashi 400-8511, Japan

[4] Computational Engineering Applications Unit, RIKEN, 2-1 Hirosawa, Wako, Saitama 351-0198, Japan

[5] J-PARC Center, Japan Atomic Energy Agency, 2-4 Shirakata, Tokai, Naka, Ibaraki 319-1195, Japan

*Corresponding author
e-mail: kom@eqchem.s.u-tokyo.ac.jp

**Water freezes below 0 °C at ambient pressure, ordinarily to ice Ih with an ABAB… hexagonal stacking sequence. However, it is also known to produce "ice Ic" nominally with an ABCABC… cubic stacking sequence under certain conditions[1], and its existence in Earth's atmosphere[2-4], or in comets[5,6] is debated. "Ice Ic", or called as cubic ice, was first identified in 1943 by König[7], who used electron microscopy to study the condensation of ice from water vapor to a cold substrate. Subsequently, many different routes to "ice Ic" have been established, such as the dissociation of gas hydrates, warming amorphous ices or annealing high-pressure ices recovered at ambient pressure, freezing of μ- or nano-confined water (see refs in [1]). Despite the numerous studies on "ice Ic", its structure has not been fully verified, because the diffraction patterns of "ice Ic" show signatures of stacking-disorder[1,8], and ideal ice Ic without stacking-disorder had not been formed until very recently[9]. Here we demonstrate a route to obtain ice Ic without stacking-disorder by degassing hydrogen from the high-pressure form of hydrogen hydrate, $C_2$, which has a host framework that is isostructural with ice Ic[10]. Surprisingly, the stacking-disorder free ice Ic is formed from $C_2$ via an intermediate amorphous or nano-crystalline form under decompression, unlike the direct transformations that occur in the cases of recently discovered ice XVI[11] from neon hydrate, or ice XVII[12] from hydrogen hydrate. The obtained ice Ic shows remarkable thermal stability until the phase transition to ice Ih at 250 K; this thermal stability originates from the lack of dislocations, which promote changes in the stacking sequence[13]. This discovery of ideal ice Ic will promote understanding of the role of stacking-disorder[14] on the physical properties of ice as a counter end-member of ice Ih.**

"Ice Ic" is known as a metastable form of ice at atmospheric pressure. The routes to synthesize "ice Ic" are increasing steadily[1], and there is a consensus that it is possibly present during the first stage in ice formation in clouds[15]. Recent computer simulations suggest that even stacking-disordered "ice Ic", called ice Isd[8], could be the stable phase for crystallites up to sizes of at least 100,000 molecules[14]. The stability of stacking-disordered ices is extremely important because of the ubiquitous nature of ice. Stacking-disordered ice can be characterized by the degree of ice "cubicity", $\chi$, which can be defined as the fraction of cubic layers if the stacking is random[1,8,16-18]. Until very recently, the highest cubicity was limited to ~80%[8,15], but it has been reported that ideal ice Ic with 100% cubicity has been obtained by annealing ice XVII[9].

From recent discoveries of polymorphs of ice XVI[11] and ice XVII[12,19], we hypothesized that ideal ice Ic could be obtained by degassing hydrogen from hydrogen hydrate, $C_2$. Five different phases in the $H_2$-$H_2O$ system have been reported to date (see refs in [20]): Among them, neutron diffraction experiments have never been conducted for the higher-pressure phases, $C_1$ and $C_2$, probably due to the technical difficulty in loading hydrogen into a pressure vessel, or compressing it to pressures in the giga-pascal range. To synthesize ideal ice Ic, decompression under low-temperature conditions for degassing is necessary, which is also not straight-forward using conventional pressure-temperature controlling systems. We have developed a 'Mito system[21]', and have overcome these technical difficulties (see details in Methods).

We started by using a mixture of $D_2O$ and $MgD_2$, which is an internal deuterium source, to synthesize hydrogen hydrate, $C_2$. After loading the mixture into a pressure-temperature controlling system, $MgD_2$ was decomposed by heating at 403 K and at ca. 0 GPa for 1 h through a nominal reaction of $MgD_2 + 3D_2O \rightarrow Mg(OD)_2 + 2D_2 + D_2O$ (at *b* in Fig. 1, the observed neutron diffraction patterns are shown in Extended Data Fig. 1). Then, the samples were cooled to room temperature (at *c* in Fig. 1) and typically compressed up to approximately 3 GPa until a $C_2$ phase was observed (at *d* in Fig. 1).

The neutron diffraction pattern for the $C_2$ phase obtained at 3.3 GPa and 300 K (at *d* in Fig. 1) was analyzed by the Rietveld method. We adopted a splitting site model for guest D atoms located at the 48$f$ site ($x$, 1/8, 1/8), and the host structure was identical to ice Ic[17] ($Fd\bar{3}m$, O at the 8$b$ site (3/8, 3/8, 3/8), D at the 32$e$ site ($x$, $x$, $x$)). The calculated diffraction pattern was in good agreement with the observed one, as shown in Fig. 2a. The refined structural parameters are listed in Extended Data Table 1.

The sample was then cooled from 300 K to 100 K at around 3 GPa (path $d\rightarrow e$). In the diffraction pattern taken at $e$ in Fig. 1, peaks from solid deuterium (phase I) appeared at around 200 K (Extended Data Fig. 2), which is consistent with the known melting curve of hydrogen[22]. This observation indicates that fluid deuterium coexisted with $C_2$ through the path from $b$ to $d$.

The $C_2$ phase persisted at pressures at least as low as 0.5 GPa on decompression at 100 K (path $e\rightarrow f$). However, surprisingly, the Bragg peaks of $C_2$ mostly disappeared at 0.2 GPa (Fig. 3). This phenomenon is totally unexpected, because the host structure of gas hydrates retains its framework in the previous cases with ice XVI[11] and XVII[12]. The sample was further decompressed to 0 GPa and evacuated using a turbo-molecular-pump. The broad peaks corresponding to ice Ic appeared at this stage. The peak disappearance of $C_2$ before the appearance of ice Ic was reproducibly observed in at least two separate neutron runs and one x-ray diffraction run for a hydrogenated sample (Extended Data Fig. 2). In the neutron diffraction pattern at 0.2 GPa, except for the Bragg peaks from $Mg(OD)_2$, only a broad peak was observed at around $d$ = 3.75 Å, which was between the peak positions of 111 of $C_2$ and that of Ice Ic (Fig. 3). This fact implies that this state does not have long-range periodicity like a normal crystal, but has only local-ordering like an amorphous or nano-crystal. Considering the observed $d$-spacing, this "amorphous-like" form would be an intermediate transition state from $C_2$ to ice Ic, which forms while hydrogen molecules are partially degassed. It is highly likely that this apparent amorphization is derived from the lattice mismatch between $C_2$ and ice Ic, originating from the relatively small cage in the host framework of the ice Ic structure.

From the x-ray diffraction run, ice Ic, which may partially include molecular hydrogen, even appeared at 0.1 GPa through the transition from the $C_2$ phase to the "amorphous-like" state, even under pressure (Extended Data Fig. 2). This also represents a difference from the previous cases of ice XVI and XVII; ice XVI is formed under evacuation[11], and hydrogen molecules can be refilled into ice XVII at an order of 10 bar of pressure[12]. It is worth noting that the partially degassed states are allowed in the cases of both ice XVI and XVII, so that the guest molecules can be continuously degassed from a fully occupied state to an empty state. The observed phase-separation behavior even under pressure in the ice $H_2$-$H_2O$ system indicates that the partially degassed $C_2$ phase would be unstable compared to the fully occupied or emptied phases, probably due to their lattice-mismatch.

The Bragg peaks in the neutron diffraction pattern for ice Ic obtained at 100 K were still broad, probably due to the small crystallite size and/or the remaining guest hydrogen molecules. The peaks of ice Ic sharpened with increasing temperature. This sharpening is dependent not only on temperature but also on time, which indicates that it is kinetic behavior.

We conducted a separate run in order to obtain a neutron diffraction pattern for the structure refinement of the ice Ic. In this run, the neutron diffraction pattern was obtained at 130 K, which is well below the temperature at which the nucleation of ice Ih occurs[23]. We confirmed that the peak width did not change in the temperature region from 130 K to 180 K, such that the peak sharpening was almost complete, even at 130 K. The obtained neutron diffraction pattern was well fitted using the ice Ic structure model[17], as shown in Fig. 2b and Extended Data Table 1. We also conducted the Rietveld analysis using $C_2$ structure model, and found that the occupancy of the D2 site was zero, within experimental error (occ(D2) = -0.001(1)). This shows that the guest hydrogen molecules are below the detectable limit at 130 K under evacuation. The peak profile around 111 peak of ice Ic has neither the feature of stacking-disorder nor the peaks from ice Ih, as shown in the diffraction pattern in the region at around $d$ = 3.9 Å, where the strongest 101 reflection of ice Ih is expected (see inset in Fig. 2b, and more detailed discussion for the peak broadening for ice Ic is described in Supplementary Information). This should be a clear indication of the presence of ideal ice Ic without stacking-disorders ($\chi$ =100%), as clear as the recent discovery of ideal ice Ic by annealing ice XVII[9].

It is also noteworthy that the ice Ic surprisingly persists up to at least 240 K until ice Ih started to appear at 250 K (Fig. 3). The temperature of 240 K corresponds to the upper limit of the reported metastable region of "ice Ic"[1]. However, in stacking-disordered ice, the cubic stacking sequence starts to change into a hexagonal stacking sequence at a much lower temperature, and the phase transition to ice Ih is completed at 240 K. The notable stability of the ice Ic would be derived from the lack of stacking-disorder. The stacking-disordered ice has more dislocations, which promote the phase transformation from "ice Ic" to ice Ih by reducing the activation energy required to change the stacking sequence[13]. This is also supported by a recent mesoscopic-size calculation[24]. The diffraction pattern observed at 250 K looks a mixture of bulk ice Ic and Ih, rather than stacking-disordered ice with many stacking faults, judging from "stackogram" reported in the literature[8,18]. At 250 K, crystal growth would be dominant, rather than crystal nucleation. Therefore, once a crystallite nucleates, it quickly grows before other crystallites nucleate, resulting in the mixture of ice Ic and Ih, rather than stacking-disordered ice. This observation also suggests a smaller number of dislocations in the ice Ic observed in this study. On the contrary, the remarkable stability of the ice Ic and the bulk mixture of ice Ic and Ih at 250 K strongly supports the conclusion that the obtained ice is not stacking-disordered, and it can therefore be called ice Ic without the need for quotation marks.

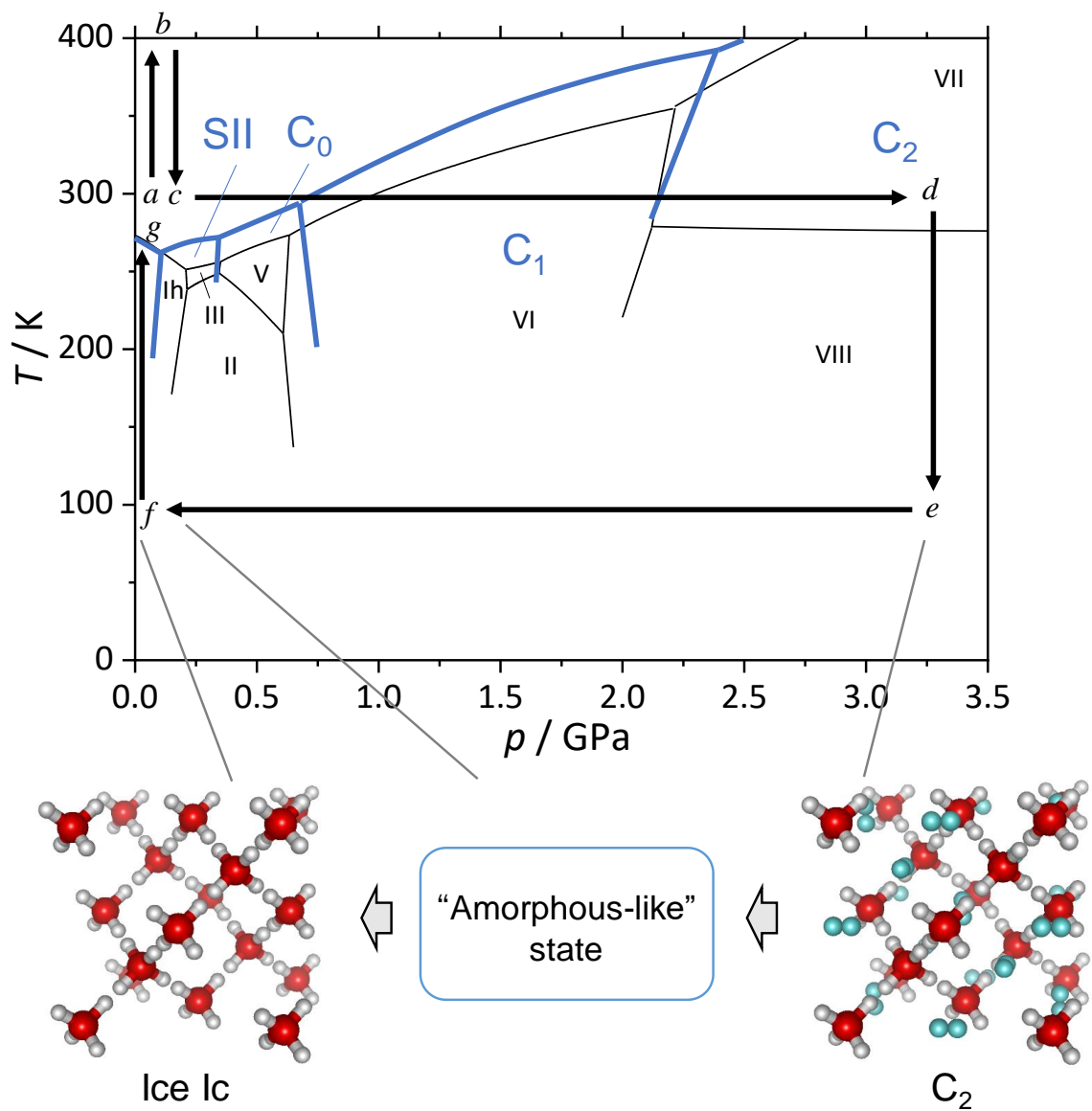

**Figure 1 | Phase diagram of hydrogen hydrate and ice with experimental paths in this study.** Phase boundaries for hydrogen hydrates and ices are drawn using thick blue lines and thin black lines, respectively. Experimental *p-T* paths are shown as black arrows in alphabetical sequence from *a* to *g*. The structural models for a high-pressure form of hydrogen hydrate, $C_2$, and ice Ic are schematically drawn with a newly found "amorphous-like" state as an intermediate transitional state from $C_2$ to ice Ic. Red, white, and blue balls in the structure model depict oxygen, hydrogen in water molecules, and hydrogen in guest molecules, respectively.

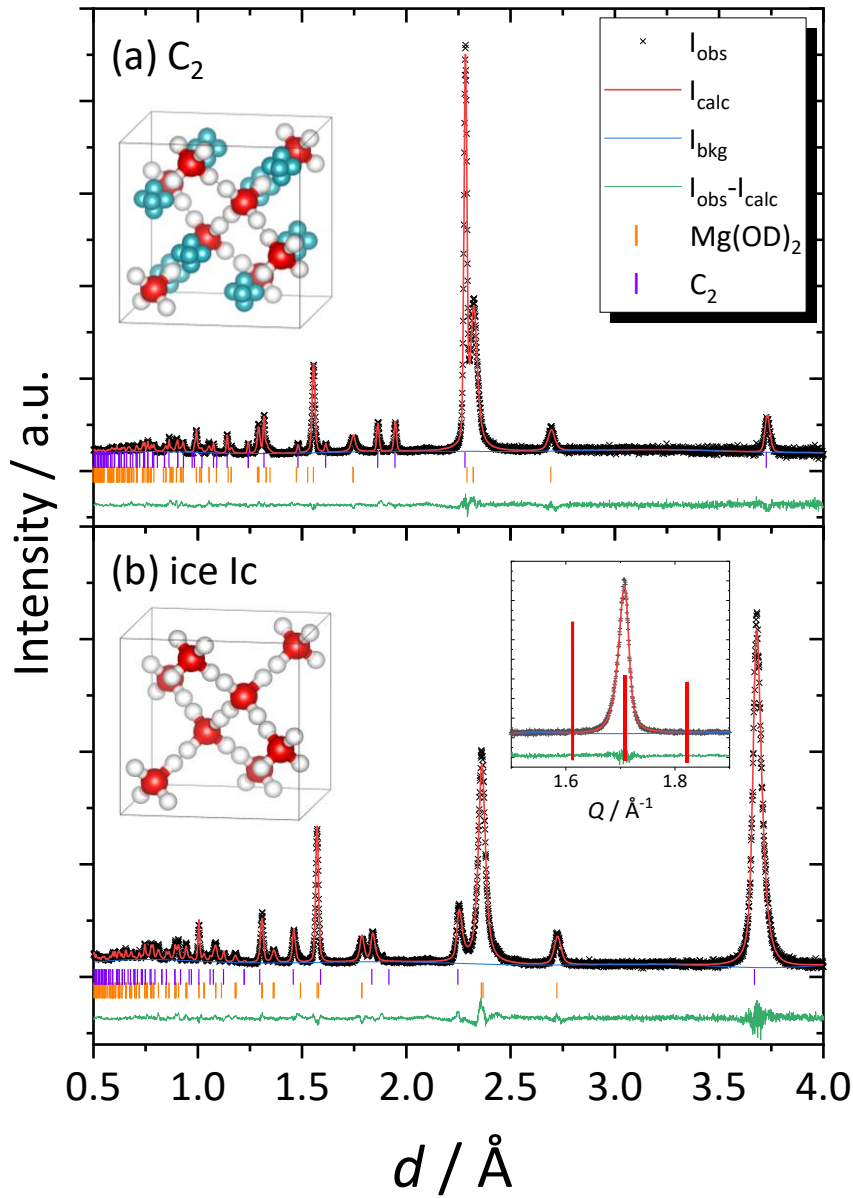

**Figure 2 | Results of Rietveld analyses for neutron diffraction patterns of (a) hydrogen hydrate, C$_2$, and (b) ice Ic.** The patterns of C$_2$ and ice Ic were obtained at 3.3 GPa and 300 K (at *d* in Fig. 1), and at 0 GPa and 130 K (in the path *f→g*). The inset diffraction pattern in (b) shows the expanded area for 111 reflections. The calculated peak positions of ice Ih are shown as red lines in the inset. Structure models for C$_2$ and ice Ic are also shown as insets in (a) and (b), respectively.

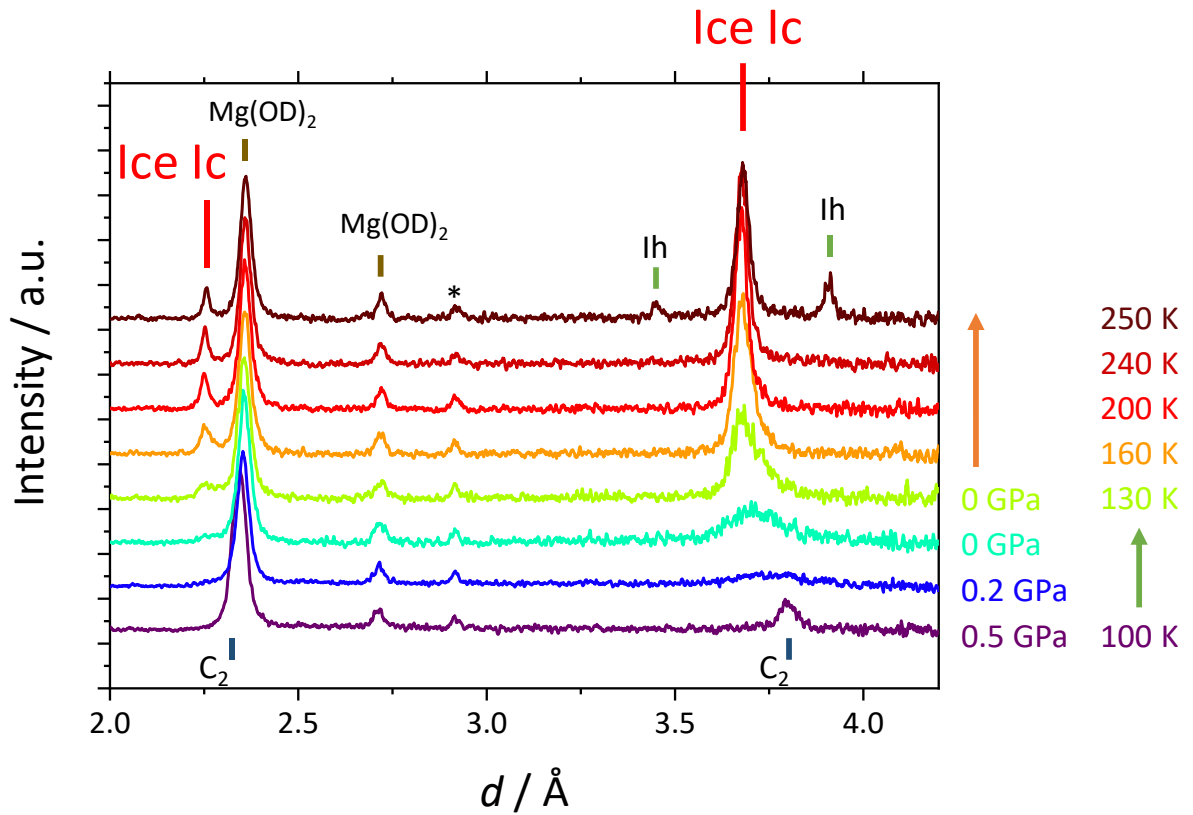

**Figure 3 | Neutron diffraction patterns with decreasing pressure at 100 K (path *e→f*) and with increasing temperature at 0 GPa (path *f→g*).** Corresponding temperatures and pressures are shown at the right side of the respective patterns. Most observed peaks are identified as $C_2$, ice Ic, $Mg(OD)_2$, or ice Ih. The peak marked by an asterisk is a parasitic peak from the high-pressure cell.

**Acknowledgments**

We are grateful to Drs. J. Abe and K. Funakoshi for their assistance with the experiments. Neutron diffraction experiments were performed through the J-PARC user programs (Nos. 2014B0187, 2015A0033, 2017A0092, 2017B0031). This study was supported by JSPS KAKENHI (Grant Numbers: 18H05224, 18H01936, 15H05829).

**Author contributions**

KK and SM conceived and designed the experiments. KK, SM, TH, ASF, RY, KY, and HK conducted the neutron diffraction experiments. KK, RY, KY, and HK conducted the x-ray diffraction experiments. FN carried out the DFT calculations and wrote the corresponding part of the manuscript. KK analyzed the data and wrote the manuscript with contributions from all authors.

## Methods

### Synthesis of MgD$_2$

MgD$_2$, used as the starting material in this study, was synthesized from reagent-grade MgH$_2$ as follows. MgH$_2$ powder (Wako pure chemical industries, Ltd.) was purchased and further ground in an agate motor to increase the surface area, after which it was placed in a copper tube with a diameter of 4 mm and a length of 40 mm. The tube was mechanically sealed and but not welded, allowing the transfer of hydrogen gas. The copper tube was inserted into a 1/4'' Inconel tube and connected in parallel to a deuterium gas cylinder and a turbo molecular pump (TMP) with 1/16'' stainless tubes and stop bulbs. The Inconel tube, including the sample copper tube, was heated to 773 K for 1 h using a tube furnace under evacuation using the TMP. Under these conditions, MgH$_2$ completely decomposed to Mg and H$_2$[25], and the degassed H$_2$ was evacuated. Then, the D$_2$ gas was introduced up to 4 MPa, and the temperature was cycled at the rate of 1 K/min between 673 K and 773 K while keeping the pressure at 4 MPa, which represents stable and unstable conditions for MgH$_2$[25], and this temperature cycle was repeated for 20 times. This 'activation' process is necessary for the reaction Mg + D$_2$ → MgD$_2$. Finally, the $p$-$T$ conditions were maintained at 673 K and 4 MPa for 3 days to complete the reaction. The recovered sample was analyzed by powder x-ray diffraction (MiniFlex-II, Rigaku) and identified to be MgD$_2$ with a trace amount of MgO. Both MgD$_2$ and MgO react with D$_2$O and produce Mg(OD)$_2$, so this small amount of MgO does not affect the conclusion.

### Neutron diffraction and $p$-$T$ control

Neutron diffraction experiments were conducted at the beamline PLANET[26] in the Material and Life Science Experiment Facility (MLF) of J-PARC, Ibaraki, Japan. Approx. 20 mg of MgD$_2$, synthesized as described above, was filled into TiZr null scattering gaskets, and D$_2$O water (99.9 %, Wako pure chemical industries, Ltd.) was dropped on the MgD$_2$ powder, resulting in the molar ratio of MgD$_2$:D$_2$O ~ 1:3. The gaskets were sandwiched between a pair of tungsten carbide anvils, and loaded by using a "hybrid Mito system", which is a modified version of an original pressure-temperature variable "Mito system[21]". The hybrid Mito system uses both flowing liquid nitrogen and a 4 K cryostat (RDK-415D, Sumitomo Heavy Industries, Ltd.), which allows us to control temperature rapidly, owing to the large latent heat of liquid nitrogen and efficient thermal insulation by zirconia and GFRP seats. The hybrid Mito system also allows us to achieve temperatures below 77 K, and reach a minimum temperature of approximately 35 K, owing to the cryostat. Another remarkable feature of the hybrid Mito system is that it affords pressure control, even at low temperature, as well as the original Mito system, which is indispensable for this study. Flexible copper cloths were attached on the support rings of the anvils, and the cloths were placed in contact with the cold head of the cryostat for thermal conduction. The accessible minimum temperature of the hybrid Mito system is ~35 K, which may be the current technical limitation due to an unavoidable influx of heat from the surrounding cell. The sample pressure was estimated from the observed lattice parameter of Mg(OD)$_2$ brucite using the equation of states[27] and the observed unit cell volume of brucite at 0 GPa, assuming the temperature derivative of the bulk modulus of brucite, $dK/dT$, was approximately 0. Although this assumption may cause some error in the pressure estimated at low temperature, we placed emphasis on avoiding unwanted Bragg peaks from additional sources of pressure. Moreover, the error would be too small to affect the conclusion. The sample position was aligned by scanning to maximize the sample scattering intensity. The Rietveld analyses were performed using the GSAS[28] with EXPGUI[29], and the crystal structure was drawn with the VESTA program[30]. Detailed procedures for data reduction and refinements are described elsewhere[31].

**X-ray diffraction**
Powder x-ray diffraction measurements using a H$_2$O (Milli Q) and MgH$_2$ (Wako pure chemical industries, Ltd.) mixture as starting materials were performed at the beamline BL-18C in the Photon Factory (KEK, Tsukuba, Japan). Samples were exposed to 0.6134 Å monochromatized synchrotron radiation, and the diffracted scattering was detected by an imaging plate (IP). Details of data reduction procedures for x-ray diffraction are described elsewhere[21]. The pressure was generated using CuBe alloy diamond-anvil cells and the temperature was controlled using a 4K GM cryostat (MiniStat, Iwatani Co.) equipped with a temperature controller (Model 335, Lakeshore). Sample pressure was estimated from the difference in the R1 line wavelengths of rubies inside and outside the sample chamber[32]. The temperature was monitored using a Si-diode sensor inserted in the cold head edge. We confirmed that the measured temperature was almost the same as that of the diamond anvils after temperature stabilization. The experimental $p$-$T$ path was basically identical to the case of neutron diffraction, as shown in Fig. 1, while the achieved pressure at path $d$ was 4.1 GPa.

One conically shaped Boehler-Almax type diamond anvil[33] with a 0.6 mm culet was placed in the direction of the detector with an opening angle of $2\theta < 40°$, whereas a conventional anvil with a 0.8 mm culet was positioned in the direction of the x-ray source. A CuBe plate with a hole of diameter 0.3 mm and an initial thickness of 0.2 mm was used as a gasket. This gasket was not subjected to pre-indentation. The load was applied by driving the piston by bellows using a He gas cylinder. The bellows allow us to control pressure at a few kbar more precisely than conventionally used membranes.

**DFT calculations**

Quantum Espresso[34] was used for the DFT calculations[35,36]. We used Perdew-Burke-Ernzerhof (so-called PBE) type nonempirical exchange-correlation functions[37] for this study. The pseudopotentials were derived using projector augmented-wave approximation[38]. The dispersion effects were taken into account using the exchange-hole dipole moment method (XDM), which calculates coefficients for polynomial of DFT-D dispersion energy[39] from the exchange-hole dipole moment calculated from simulated electron wave function[40,41]. XDM damping function parameters are taken from [42]. The enthalpies of four possible configurations for the ordered form of ice Ic were calculated within a unit cell with a kinetic energy cutoff of 70 Ry and a Brillouin zone k mesh of $8 \times 8 \times 8$. The cell and atomic parameters were optimized using BFGS quasi-Newtonian methods at atmospheric pressure.

**Extended Data Table 1 | Experimental details and refined crystallographic data for C$_2$ and ice Ic.**

|  | C$_2$ | Ice Ic |
|---|---|---|
| Chemical formula | D$_2$·D$_2$O | D$_2$O |
| Molecular weight | 40.055 | 20.028 |
| Crystal system | Cubic | Cubic |
| Space group | $Fd\bar{3}m$ | $Fd\bar{3}m$ |
| Temperature (K) | 300 | 100 |
| Pressure (GPa) | 3.3 | 0 |
| a (Å) | 6.45307(8) | 6.3560(2) |
| V (Å3) | 268.719(11) | 256.77(3) |
| Z | 8 | 8 |
| Radiation type | Spallation neutron | |
| Diffractometer | PLANET (BL11), MLF, J-PARC | |
| Specimen mounting | Pressure-temperature controlling system (the hybrid MITO system) | |
| $R_p$ | 0.0379 | 0.0532 |
| $R_{wp}$ | 0.0425 | 0.0605 |
| $R(F^2)$ | 0.114 | 0.0772 |
| $\chi^2$ | 1.975 | 2.556 |
| No. of data points | 3525 | 3525 |
| No. of parameters | 29 | 24 |
|  |  |  |
| $U_{iso}$(O) | 0.0113(7) | 0.0181(7) |
| $x$(D1*) | 0.4607(2) | 0.4675(3) |
| occ(D1*) | 0.5 (fixed) | 0.5 (fixed) |
| $U_{iso}$(D1*) | 0.0250(9) | 0.0239(7) |
| $x$(D2**) | 0.0636(5) | - |
| occ(D2**) | 0.310(1) | - |
| $U_{iso}$(D2**) | 0.0250(9) ($=U_{iso}$(D1)) | - |

*D1 belongs to a water molecule in the host structure.
**D2 belongs to the guest deuterium molecule. $U_{iso}$(D2) is constrained to be the same value as $U_{iso}$(D1), because of the severe correlation between atomic coordinates and occupancies.

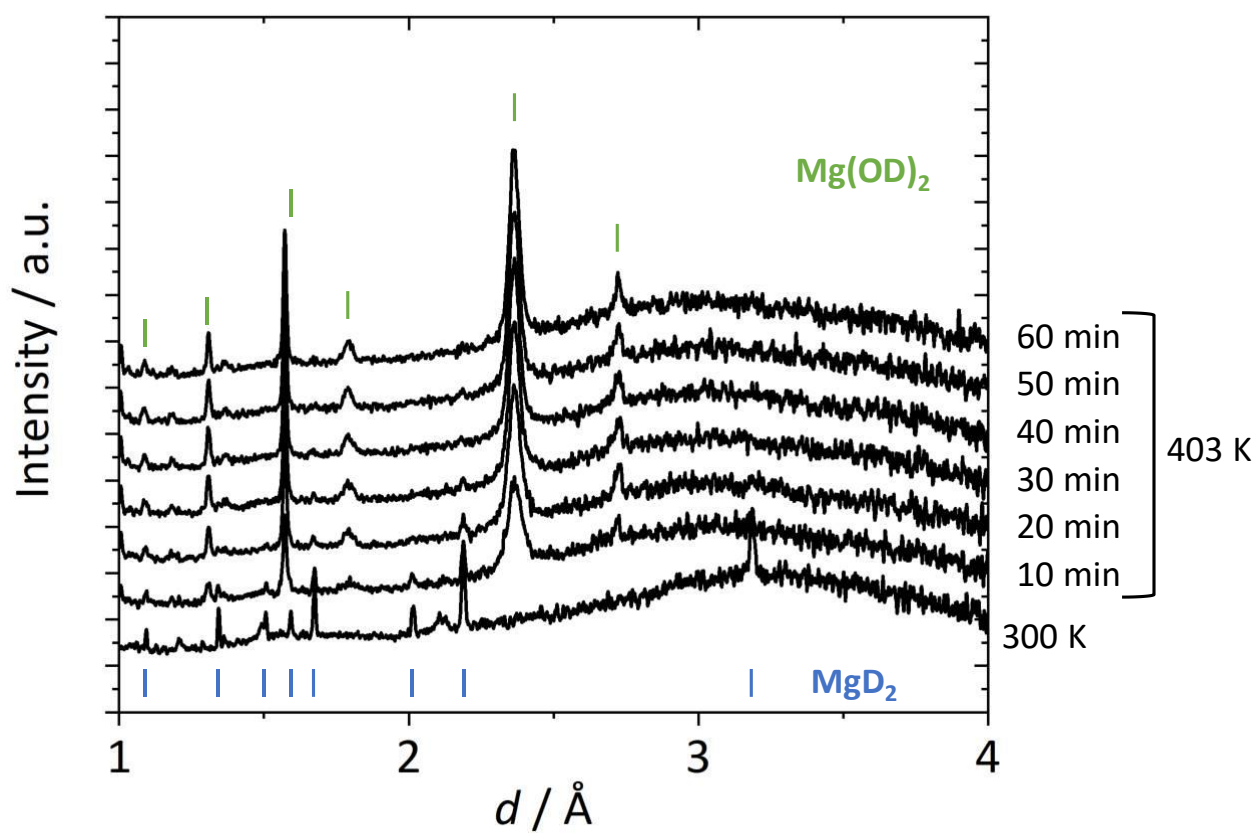

**Extended Data Figure 1 | Neutron diffraction patterns for starting materials taken at 300 K (bottom, at *a* in Fig. 1), and time-resolved patterns at 403 K (at *b* in Fig. 1) for each 10 min.** The obtained Bragg peaks are indexed as $MgD_2$ (blue tick marks) or $Mg(OD)_2$ (green tick marks). A broad peak at around $d \sim 3$ Å in the pattern at 300 K originated from liquid $D_2O$, and the broad peak shifted to lower *d*-spacing at 403 K, which would show the existence of fluid $D_2$ in the sample chamber.

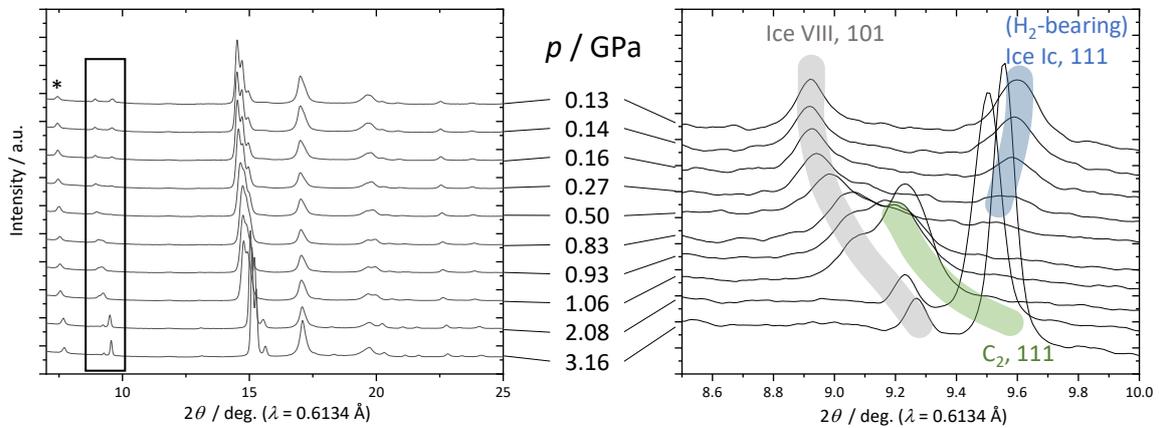

**Extended Data Figure 2 | (a) X-ray diffraction patterns at 100 K with decreasing pressure (path *e→f*) and (b)** an enlarged $2\theta$ region from 8.5° to 10°, corresponding to the region shown in the black box in the figure on the left. Thick blue, green, and gray lines in the figure on the right schematically show the peak positions of 111 of ice Ic and $C_2$, and 101 of ice VIII, respectively. Because the $MgH_2$:$H_2O$ ratio in the starting material of this x-ray diffraction run was not sufficiently high to make pure $C_2$, ice VIII remained in this run. At 0.50 GPa, 111 peaks of both $C_2$ and ice Ic mostly disappeared, indicating the intermediate amorphous-like state at that pressure. The asterisk in the left figure denotes a scattering from Mylar® (polyester) film, which is used for a window material of the vacuum chamber.

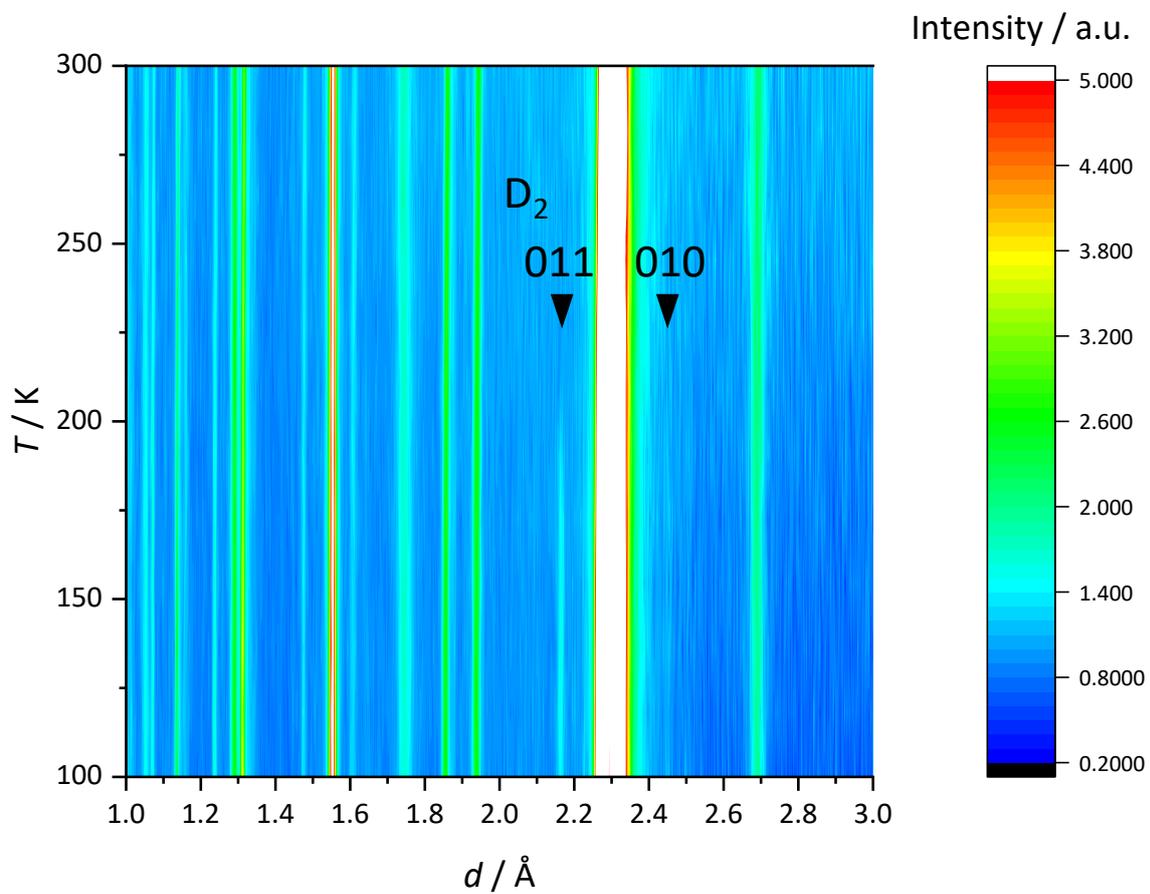

**Extended Data Figure 3 | Sequential neutron diffraction patterns with decreasing temperature (path *d→e*).** Two 011 and 010 peaks from solid $D_2$ (Phase I) are clearly seen, as shown by black triangles.

# Supplementary Information

**On the peak broadening of ice Ic**

The peak width of the ice Ic obtained in this study ($\Delta d/d \sim 1.2\%$) are significantly larger than the instrument resolution ($\Delta d/d \sim 0.6\%$)[26], as shown in Supplementary Fig. S1. This broadening cannot be the result of stacking-disorder, because the profile changes caused by the stacking-disorder are not simply broadening, but more complicated[8,18]. Considering the result that the peak width did not change up to 240 K once the peak sharpening converged, it is unlikely that the peak broadening of ice Ic could originate only from small crystalline size. Other broadening factors such as microstrain induced by lattice distortions should be taken into account (see also Supplementary Fig. S1). We assume the lattice distortion could be caused by proton ordering, and then conducted the DFT calculations for the ordered form of ice Ic in order to estimate the degree of lattice distortion. There are four possible symmetrically non-equivalent configurations for ice Ic with unit cell volumes identical to that of the disordered ice Ic, having the space groups of $Pna2_1$, $I4_1md$, $P4_1$, and $P4_12_12$[43]. The optimized unit cell parameters show deviations from the cubic lattice (Supplementary Table S1). The deviation, which can be defined as the ratio of cell parameters between the characteristic axis (= the most deviated axis from cubic symmetry) and the other axes, ranges from 0.12 % for the $P4_1$ model to 1.16 % for the $P4_12_12$ model. The degree of deviation is of an order similar to the peak broadening of 0.6% (1.2 % − 0.6 %) added to the instrument resolution. Although the obtained ice Ic does not have the long-range ordering for protons, short-range ordering could result in the lattice distortion, which contributes to the peak broadening.

**Supplementary Table S1 | Results of DFT calculations for four symmetrically non-equivalent ordered forms of ice Ic.**

| Order model | 1 | 2 | 3 | 4 |
| --- | --- | --- | --- | --- |
| Crystal system | Orthorhombic[*] | Tetragonal | Tetragonal | Tetragonal |
| Space Group | $Pna2_1$[*] | $I4_1md$ | $P4_1$ | $P4_12_12$ |
| Dipole Moment (Debye) | 5.64 | 7.96 | 3.95 | 7.65E-05 |
| $a$ | 6.1953 | 6.1711 | 6.1806 | 6.1352 |
| $b$ | 6.1953 | 6.2007 | 6.1878 | 6.2070 |
| $c$ | 6.1547 | 6.1711 | 6.1806 | 6.2070 |
| $\alpha$ | 90.0000 | 90.0000 | 90.0000 | 90.0000 |
| $\beta$ | 90.0000 | 90.0000 | 90.0000 | 90.0000 |
| $\gamma$ | 90.0242 | 90.0000 | 90.0000 | 90.0000 |
| Charac. axis /Cubic axis[**] | 0.9935 | 1.0048 | 1.0012 | 0.9884 |
| $H$(eV)/8H$_2$O | -4799.3535 | -4799.3755 | -4799.3358 | -4799.3249 |
| $\Delta H$(eV)/8H$_2$O | 0.0219281 | 0 | 0.0396700 | 0.0505637 |

[*]DFT calculation was started from the orthorhombic $Pna2_1$ structure, but the symmetry of the optimized structure could be reduced from orthorhombic to monoclinic.

[**]Ratio of cell parameters between the characteristic axis and cubic axis. The characteristic axis shows the greatest deviation from the cubic axis, as shown by shades.

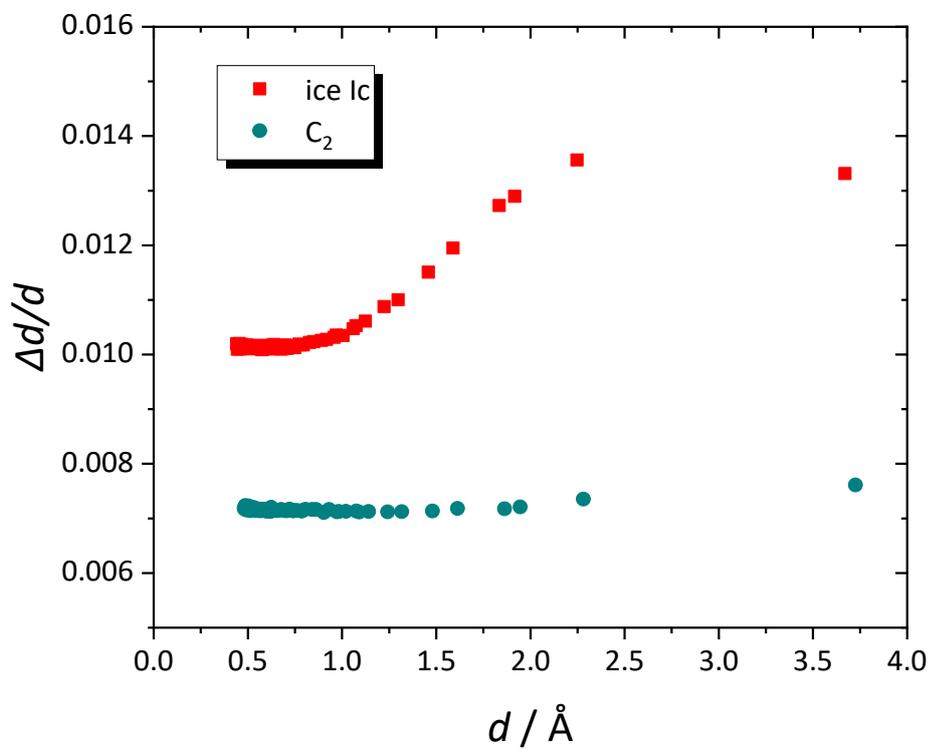

**Supplementary Figure S1** | Full width at half maximum divided by *d*-spacing (Δ*d*/*d*) as a function of *d*-spacing for $C_2$ and ice Ic.